\documentstyle [12pt,epsfig] {article}
\begin{document}

\centerline{\bf CONSTITUENT QUARK-BASED LINEAR $\sigma$ MODEL (L$\sigma$M)}
\centerline{\bf QUARK AND SCALAR MESONS, VECTOR MESON DOMINANCE}
\vskip 0.3cm

\centerline{Michael D. Scadron\footnote{E-mail address:
scadron@physics.arizona.edu}}
\centerline{\sl Physics Department, University of Arizona, Tucson,
Arizona 85721,USA}
\centerline{Miroslav Nagy\footnote{E-mail address: fyzinami@savba.sk}}
\centerline{\sl Institute of Physics, Slovak Academy of Sciences, 
845 11 Bratislava, Slovakia}

\vskip 0.7cm
\begin{abstract}
After describing the SU(2) linear sigma model (L$\sigma$M), we
dynamically generate it using the B.W. Lee null tadpole sum
(characterizing the true vacuum) together with the dimensional
regularization lemma. Next we generate the chiral-limiting (CL)
nonstrange and strange constituent quark masses ${\hat m}=325.7$ MeV;
$m_s=486$ MeV away from the CL. Finally, we study vector meson
dominance (VMD) and the pion, kaon charge radii and the loop-order
$\rho\to\pi\gamma$, $\pi^0\to\gamma\gamma$ amplitudes in the quark
model. Lastly, we verify this procedure using tree-order VMD graphs. 
\end{abstract}
\vskip0.7truecm

\centerline{\bf 1$~$ Introduction}
\vskip0.7truecm

It is well-known that hadron low-energy interactions are quite
well-described by the effective chiral Lagrangians (ECL). The ECL
represents the Lagrange form of an approximate chiral-symmetry
realization which is also the case for quantum chromodynamics (QCD).
The ECL method allows a deeper insight into the current algebra
results, the low-energy theorems as well as the hypotheses of the
vector meson dominance (VMD) and partial conservation of axial current
(PCAC). The attractive peculiarity of the ECL formalism is the
possibility to work directly with the observable (physical) particles
and describe observable quantities. The study of the ECL properties is
an actual problem in modern particle physics.

The derivation of ECL from "the first principles" in QCD is a problem
not solved so far. Its solution needs joint description of such effects
as bosonization, spontaneous chiral symmetry breaking (SBCS) and
confinement. Unfortunately at the contemporary level of development of
quantum field theory, the joint description "from first principles" of
these complicated nonperturbative effects has not been achieved. That
is why for the study of the ECL properties one frequently uses the
effective quark models based on QCD.

The most convenient quark model, describing at the quantum field
theory level such non-perturbative effects as SBCS and bosonization, is
the quark-based L$\sigma$M. The conversion of current quarks into
constituent ones (i.e. SBCS), the elimination of quark degrees of
freedom and the transition to low-lying meson states can be performed
at the quantum field theory level.

This article is organized as follows. After introduction in Sec. 2 we
first use four chiral couplings $g$, $g'$, $\lambda$, $m_q$ to describe
the quark-based L$\sigma$M interacting Lagrangian density. Then we
follow B.W. Lee and chracterize the true (vanishing) vacuum via the
null tadpole sum of $u$, $d$ quark loops plus the pion loop plus the
sigma meson loop. Even though these loops are all quadratically
divergent, their sum must vanish independent of any renormalization
scheme. We next solve this vanishing sum via dimensional analysis and
then via dimensional regularization. Next in Sec. 3 we invoke a
once-subtracted dispersion relation for the pion decay constant. The
result is $f_{\pi}$ varies from $q^2=0$ to $m_{\pi}^2$ by only about
3\%. Also the nonstrange constituent quark mass varies from the chiral
limit (CL) to its on-shell value (to about 337.4 MeV) by 3.6\%. Finally
in Sec. 4 we extend the analysis to strange quarks, $SU(3)$ scalar
mesons and to meson charge radii, using both the above L$\sigma$M and
independetly VMD.

\vskip0.7truecm
\centerline{\bf 2$~$ $SU(2)$ L$\sigma$M Lagrangian density}
\vskip0.7truecm

First we display the interacting part of the $SU(2)$ L$\sigma$M
Lagrangian density 

\begin{equation}
{\it L}^{\rm int}_{L\sigma M}= g{\bar \psi}(\sigma + i\gamma_5{\vec
\tau}.{\vec \pi})\psi + g'\sigma(\sigma^2 + \pi^2) -
\frac{\lambda}{4}(\sigma^2 + \pi^2)^2 - f_{\pi}g{\bar\psi}\psi
\label{L_int}
\end{equation}
\noindent for approximately $f_{\pi}\approx 93$ MeV. The corresponding
fermion and meson Goldberger-Treiman relations (GTRs) are
\begin{equation}
g=\frac{m_q}{f_{\pi}}  \label{gmfpi}
\end{equation}
\begin{equation}
g'=\lambda f_{\pi} = m^2_{\sigma}/2f_{\pi}.  \label{lambda}
\end{equation}
The original L$\sigma$M \cite{GML} involving nucleons as the fermion
fields in Eq.(\ref{L_int}) required spontaneous symmetry breaking (SSB)
in order that chiral symmetry holds.

Alternatively, if the fermion fields in Eq.(\ref{L_int}) correspond to
constituent $u$ and $d$ quarks, SSB is replaced by dynamical symmetry
breaking (DSB) \cite{DS}, characterized by the true vacuum determined
by the B.W. Lee null tadpole sum \cite{BWL} of Fig.1: 

\begin{figure}[ht]
\centering
\epsfig{file=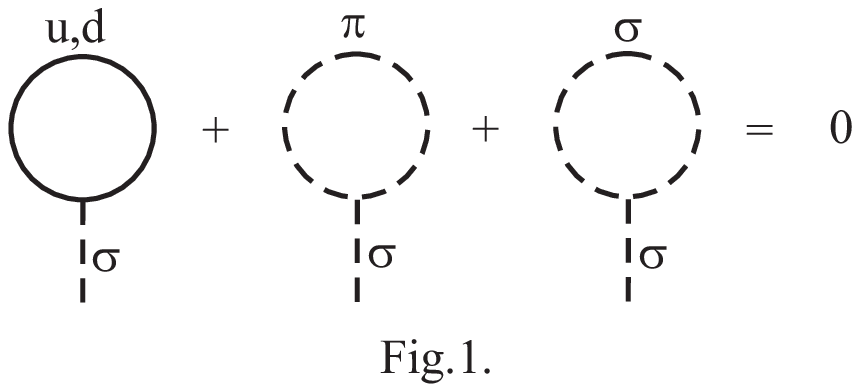, width=7cm}
\end{figure}

\noindent Here the average quark mass ${\hat m} = {1\over 2}(m_u+m_d)$
is generated by the GTR ${\hat m}=f_{\pi}g$. The dimensionless
meson-quark coupling $g$ and quartic meson coupling $\lambda$ obey the
GTRs of Eqs.(\ref{gmfpi},\ref{lambda}). The massive cubic meson
coupling $g'$ is $m^2_{\sigma}/2f_{\pi}$ in Eq.(\ref{lambda}).

In order to ''solve'' the null tadpole sum condition of Fig.1. \cite{DS}
we characterize these quadratic divergent tadpole integrals using
dimensional analyses:
\begin{equation}
\int\frac{d^4p}{p^2-m^2}~\propto~m^2,~~~ \int\frac{d^4p}{p^2}=0,~~~
\int\frac{d^4p}{p^2-m_{\sigma}^2}~\propto~m_{\sigma}^2.  \label{diman}
\end{equation}
Then we invoke the GTRs Eqs.(\ref{gmfpi},\ref{lambda}) and also the
$\sigma-\sigma-\sigma$ combinatoric factor of 3 to express Fig.1. in
the chiral limit (CL) as \cite{DS}
\begin{equation}
N_c(2m_q)^4 = 3m^4_{\sigma},  \label{N_c}
\end{equation}
where $N_c$ is the color number of quarks (presumably $N_c=3$).

Next we invoke the dimensional regularization lemma \cite{DS}

\begin{equation}
\lim_{l\rightarrow 2}\int d^{2l}p\left[\frac{m^2}{(p^2-m^2)^2} -
\frac{1}{p^2-m^2} \right] = \lim_{l\to
2}\frac{im^{2l-2}}{(4\pi)^l}[\Gamma(2-l) +\Gamma(1-l)]=
-\frac{im^2}{(4\pi)^2}.  \label{dim-reg}
\end{equation}
This dimensional regularization condition follows because of the gamma
function identity 
\begin{equation}
\Gamma(2-l) + \Gamma(1-l) = \frac{\Gamma(3-l)}{1-l}\to -1~~{\rm as}~~
l\to 2.   \label{Gamma}
\end{equation}
In fact, this dimensional regularization lemma Eq.(\ref{dim-reg}) also
is valid by involving a partial -fraction identity combined with the
massless quadratic tadpole $\int{d^4p}p^{-2}=0$ \cite{MDS}.

Finally one expresses the squared $\sigma$ mass via the bubble plus
tadpole graphs of Fig.2: 

\begin{figure}[ht]
\centering
\epsfig{file=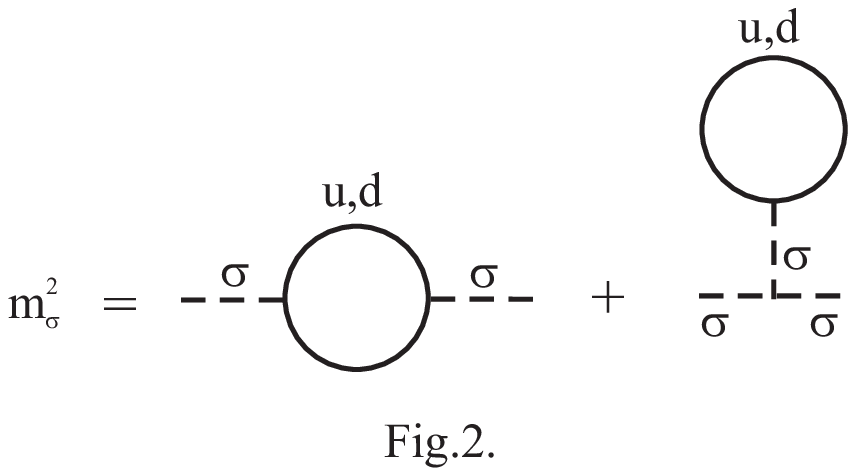, width=7cm}
\end{figure}

Since the bubble graph is logarithmically divergent and the tadpole
graph of Fig.2 is quadratic divergent, invoking the
dimensional regularization lemma of Eq.(\ref{dim-reg}), Fig.2 implies
\begin{equation}
m^2_{\sigma} = N_c g^2 \frac{m^2_q}{\pi^2}.   \label{m-sigma}
\end{equation}
Simultaneously solving Eqs.(\ref{N_c})(\ref{dim-reg})(\ref{m-sigma}),
one is led to in the CL
\begin{equation}
N_c=3,~~~m_{\sigma} = 2m_q~({\rm NJL}),   \label{NJL}
\end{equation}
\begin{equation}
g=\frac{2\pi}{\sqrt3} \approx 3.6276 ~~~(Z=0~~{\rm cc}).  \label{gCL}
\end{equation}
Eq.(\ref{NJL}) is the NJL condition \cite{NJL} and Eq.(\ref{gCL}) is
the $Z=0$ compositeness condition \cite{AS}, the latter also following
from the infrared limit of QCD \cite{ES}. These conditions
Eqs.(\ref{NJL})(\ref{gCL}) hold for the quark-level SU(2) L$\sigma$M
\cite{DS} without first referring to the nonlinear NJL model. Once the
above B.W. Lee null tadpole condition is satisfied (leading to
Eqs.(\ref{NJL})(\ref{gCL})), we suggest no further renormalization is
required. 
\eject
\vskip0.7truecm
\centerline{\bf 3$~$ Dispersion relation for the pion decay constant}
\vskip0.7truecm

Next we dynamically generate the SU(2) and SU(3) constituent quark
masses by first invoking a once-subtracted dispersion relation for the
pion decay constant \cite{CS}\cite{NSH}:
\begin{equation}
\frac{f_{\pi}}{f_{\pi}^{CL}} - 1 = \frac{m^2_{\pi}}{8\pi^2 f^2_{\pi}}
\left(1 + \frac{m^2_{\pi}}{10{\hat m}^2} \right) \approx 2.946~\%.
\label{fpiCL} 
\end{equation}
Note that the order $m^2_{\pi}$ term 2.88\% \cite{CS} increases to
2.95\% including $m^4_{\pi}$ corrections \cite{NSH}.

Now invoking the presumably observed pion decay constant \cite{PDG}
using Eq.(\ref{fpiCL}):
\begin{equation}
f_{\pi}\approx 92.42~{\rm MeV}~\&~ f_{\pi}^{CL}=
\frac{f_{\pi}}{1.02946}\approx 89.775~{\rm
MeV}.   \label{f_pi}
\end{equation}
Then the quark-level GTR in the CL is (via Eq.(\ref{gCL}))
\begin{equation}
{\hat m}^{CL} = f_{\pi}^{CL} g = 325.7~{\rm MeV},  \label{mCL}
\end{equation}
near $m_N/3\approx 313$ MeV as expected. Away from the CL the
nonstrange constituent quark mass can be estimated via the
proton magnetic dipole moment as \cite{BLP}\cite{MDS1} 
\begin{equation}
{\hat m} \approx \frac{m_N}{\mu_p}\approx \frac{938.9}{2.7928}~{\rm
MeV} \approx 336.2~{\rm MeV},   \label{mup}
\end{equation}
just 3.2\% greater than the CL quark mass in Eq.(\ref{mCL}). Note that
Eq.(\ref{mup}) is near the GTR quark mass away from the CL:
\begin{equation}
m_q = f_{\pi} g\approx 93~{\rm MeV}\times \frac{2\pi}{\sqrt3}\approx
337.4~{\rm MeV},  \label{m_q}
\end{equation}
where we again have invoked the L$\sigma$M \cite{DS} or the $Z=0$ cc  
$\Longrightarrow$ $g=2\pi/\sqrt3$ \cite{AS}.

\vskip0.7truecm
\centerline{\bf 4$~$ The extension of the analysis to strange quarks}
\vskip0.7truecm

Because of the continued consistency with these SU(2) relations, we
next extend this theory to SU(3). To introduce the strange constituent
quark, we extend the nonstrange GTR ${\hat m}=f_{\pi}g=(1/2)(m_u+m_d)$
to strange quarks obeying $f_Kg=(1/2)(m_s+{\hat m})$ with ratio
(independent of $g$):
\begin{equation}
\frac{m_s}{\hat m} = \frac{2f_K}{f_{\pi}}-1\approx 1.44   \label{msm}
\end{equation}
because \cite{PDG} $f_K/f_{\pi}\approx 1.22$. Then invoking ${\hat
m}\approx 337.4$ MeV from Eq.(\ref{m_q}), the strange quark mass in
Eq.(\ref{msm}) is
\begin{equation}
m_s\approx 1.44~{\hat m}\approx 485.9~{\rm MeV},   \label{m_s}
\end{equation}
only 5\% less than the original estimate \cite{BLP,MDS1} $m_s\approx
510$ MeV. Another test of this $m_s$ mass scale is via the kaon GTR
$m_s = 2f_Kg -{\hat m}=(823.2 -337.4)$ MeV = 485.8 MeV, in excellent
agreement with Eq.(\ref{m_s}) - but now depending on the
$g=2\pi/{\sqrt3}$ scale.

Note that the analogue NJL (or L$\sigma$M) condition in the CL implies
\begin{equation}
m_{f_0}(980) = 2m_s\approx 972~{\rm MeV},   \label{m_f_0}
\end{equation}
then compatible because data says \cite{PDG} $m_{f_0}(980)=980\pm 10$
MeV, suggesting that the $f_0(980)$ is scalar mainly ${\bar s}s$. This
is consistent with the nonstrange $a_1(1260)$ (having $a_1\to\sigma\pi$
seen, but $a_1\to f_0(980)\pi$ not seen \cite{PDG}).

Further note that the kappa scalar obeys the NJL condition
\begin{equation}
m_{\kappa} = 2[m_s{\hat m}]^{1/2}\approx 809~{\rm MeV}  \label{kappa}
\end{equation}
for ${\hat m}\approx 337$ MeV, $m_s\approx 486$ MeV, is then also
compatible with the observed E 791 data \cite{E791} $m_{\kappa}\approx
797\pm 19$ MeV. 

Lastly we estimate the nonstrange NJL - L$\sigma$M scalar mass in the
CL as
\begin{equation}
m_{\sigma}^{CL} = 2{\hat m}^{CL}\approx 651.4~{\rm MeV},  \label{msCL}
\end{equation}
invoking the nonstrange CL quark mass 325.7 MeV as obtained in
Eq.(\ref{mCL}) above. In fact Eq.(\ref{msCL}) implies the L$\sigma$M
mass (squared) away from the CL:
\begin{equation}
m^2_{\sigma} - m^2_{\pi} = (m_{\sigma}^{CL})^2~~~{\rm
or}~~~m_{\sigma}\approx 666~{\rm MeV}   \label{ms-mpi}
\end{equation}
very near the model independent \cite{SKN} $m_{\sigma}\approx 665$ MeV
based on a coupled-channel $\pi\pi\to\pi\pi,~K{\bar K}$ dispersion
analysis. An analogous coupled-channel analysis was earlier used to
estimate a kappa scalar mass in the 730 - 800 MeV region \cite{EvanB}.

In fact a still earlier infinite momentum frame [IMF] approach
\cite{MDS2} estimates the quadratic meson mass SU(3) relations
\begin{equation}
m^2_K - m^2_{\pi} = m^2_{K^*} - m^2_{\rho} = m^2_{\Phi} - m^2_{K^*}
\approx 0.22~{\rm GeV}^2.   \label{mKmpimK}
\end{equation}
These ${\bar q}q$ meson $\Delta S=1$ mass splittings are about 1/2 the
$qqq$ baryon mass splittings
\begin{equation}
{\Delta}m^2_{qqq}({\rm octet}) \approx m_{\Sigma\Lambda}^2 - m^2_N
\approx m^2_{\Xi} - m_{\Sigma\Lambda}^2\approx 0.43~{\rm GeV}^2,
\label{octet} 
\end{equation}
\begin{equation}
{\Delta}m^2_{qqq}({\rm decuplet})\approx m^2_{\Sigma^*} -
m^2_{\Delta}\approx m^2_{\Xi^*} - m^2_{\Sigma^*}\approx m^2_{\Omega} -
m^2_{\Xi^*} \approx 0.43~{\rm GeV}^2   \label{decup} 
\end{equation}
because there are two ${\Delta}S=1$ transitions for $qqq$ baryons as
opposed to ${\bar q}q$ mesons \cite{MDS2}.

Finally we study VMD. Pions and kaons are tightly bound, so the CL
${\bar q}q$ pion charge radius is
\begin{equation}
r^{CL}_{\pi} = \frac{{\hbar}c}{{\hat m}^{CL}}= \frac{197.3~{\rm
MeV}}{325.7~{\rm MeV}} {\rm fm}\approx 0.606~{\rm fm}, \label{rCL}
\end{equation}
\begin{equation}
r^{VMD}_{\pi} = \frac{{\sqrt6}{\hbar}c}{m_{\rho}}\approx 0.623~{\rm
fm},  \label{rVMD} 
\end{equation}
\begin{equation}
r^{\rm exp}_{\pi} = 0.672\pm 0.008~{\rm fm}.  \label{rexp}
\end{equation}
The above pattern also holds for kaons:
\begin{equation}
r^{CL}_K=\frac{2{\hbar}c}{m_s+{\hat m}\vert_{CL}}\approx 0.497~{\rm
fm}, \label{rKCL}
\end{equation}
\begin{equation}
r^{VMD}_K = \frac{{\sqrt6}{\hbar}c}{m_{K^*}}\approx 0.541~{\rm fm},
\label{rKVMD} 
\end{equation}
\begin{equation}
r^{\rm exp}_K = 0.560\pm 0.031~{\rm fm}.  \label{rexpK}
\end{equation}

Such tight ${\bar q}q$ binding for pions and kaons should be extended to
loosely bound $qqq$ baryons such as protons \cite{MSW,SRKB}:
\begin{equation}
R_p\approx [1+\sin 30^0]r_{\pi}\approx 0.9~{\rm fm},  \label{r_p}
\end{equation}
near data \cite{PDG}
\begin{equation}
R_p=0.870\pm 0.008~{\rm fm}.  \label{rpdata}
\end{equation}

Also note the PVV quark triangle amplitude magnitude for
$\rho\to\pi\gamma$ decay (weighted by a Levi-Civita factor) for color
number $N_c=3$
\begin{equation}
\vert F_{\rho\pi\gamma}\vert = \frac{eg_{\rho}}{8\pi^2f_{\pi}}\approx
0.206~{\rm GeV}^{-1},  \label{Frpg}
\end{equation}
as obtained via the quark triangle graph for $g_{\rho}\approx 4.97$ as 
found below. This amplitude magnitude is near data \cite{PDG}
\begin{equation}
\vert F_{\rho\pi\gamma}\vert^{\rm exp} = 0.222\pm 0.012~{\rm GeV}^{-1}.
\label{Frpgex} 
\end{equation}
Similarly the $\pi^0\gamma\gamma$ quark triangle magnitude is
\begin{equation}
\vert F_{\pi^0 2\gamma}\vert = \frac{e^2}{4\pi^2f_{\pi}}\approx 0.0251~
{\rm GeV}^{-1},  \label{Fp2g}
\end{equation}
also near data \cite{PDG}
\begin{equation}
\vert F_{\pi^0 2\gamma}\vert^{\rm exp} = 0.0252\pm 0.0009~{\rm
GeV}^{-1}.  \label{Fp2ge} 
\end{equation}

Note that VMD tree graphs require
\begin{equation}
F_{\rho\pi\gamma}\frac{e}{g_{\rho}} = \frac{F_{\pi^0 2\gamma}}{2}
\approx 0.01256~{\rm GeV}^{-1}  \label{FF}
\end{equation}
as first noted by ref.\cite{GMSW}, also near
$F_{\rho\pi\gamma}e/g_{\rho}$ data at $0.0140\pm 0.0007~{\rm
GeV}^{-1}$.  

To point out the close link between VMD and the quark level L$\sigma$M
\cite{DS,SRKB}, we first note that the L$\sigma$M requires
$g_{\rho\pi\pi}={\sqrt3}g=2\pi\approx 6.28$ as obtained by many authors
\cite{DS,LHC}. The $\rho\pi\pi$ decay rate is for $q=364$ MeV
\begin{equation}
\Gamma_{\rho\pi\pi} = \frac{g^2_{\rho\pi\pi}}{6\pi
m^2_{\rho}}q^3\approx 150.3~{\rm MeV} \Longrightarrow g_{\rho\pi\pi}
\approx 5.95,  \label{Grpp}
\end{equation}
while the smaller $\rho e^+e^-$ rate is via VMD \cite{PDG}
\begin{equation}
\Gamma_{\rho ee} = \frac{e^4m_{\rho}}{12\pi g^2_{\rho}}\approx
7.02~{\rm keV} \Longrightarrow g_{\rho}\approx 4.96.  \label{Gree}
\end{equation}
VMD universality suggests \cite{JJS} $g_{\rho\pi\pi}\approx g_{\rho}$,
which is extended in the quark-level L$\sigma$M via the $\pi - \sigma -
\pi$ meson loop \cite{BRS}:
\begin{equation}
g_{\rho\pi\pi} = g_{\rho} + \frac{1}{6}g_{\rho\pi\pi} \Longrightarrow
\frac{g_{\rho\pi\pi}}{g_{\rho}} = \frac{6}{5},   \label{6/5}
\end{equation}
very close to the data ratio in Eqs.(\ref{Grpp})(\ref{Gree}). The 1/6
factor in Eq.(\ref{6/5}) corresponds to $\lambda/(16\pi^2)$ with
$\lambda = 2g^2$ and $g=2\pi/\sqrt3$ as obtained in the L$\sigma$M
relation Eq.(\ref{gCL}) above. 

In this paper we have first reviewed the quark-level SU(2) L$\sigma$M
and then dynamically generated it. Next we obtained the CL pion decay
constant and found the nonstrange and strange constituent CL quark
masses. Then we found the pion and kaon charge radii and the quark loop
values for the $\rho\to\pi\gamma$ and $\pi^0\to\gamma\gamma$ amplitudes
- all fitting data without introducing arbitrary parameters.

\vskip0.5truecm

\noindent {\bf Acknowledgement:} This work is in part supported by the
Slovak Agency for Science, Grant 2/3105/23.

\eject
\end{document}